\documentclass[twocolumn,showpacs,preprintnumbers,amsmath,amssymb,superscriptaddress,aps,prb]{revtex4-1}

\usepackage{epsfig}

\usepackage{graphicx,latexsym,xcolor}

\usepackage{tikz}
\usepackage{amsmath}

\begin{document}

\title{Topological self-organization of strongly interacting particles
\\}
\author{Ioannis Kleftogiannis}
\affiliation{Physics Division, National Center for Theoretical Sciences, Hsinchu 30013, Taiwan}
\author{Ilias Amanatidis}
\affiliation{Department of Physics, Ben-Gurion University of the Negev, Beer-Sheva 84105, Israel}

\date{\today}
\begin{abstract}
We investigate the self-organization of strongly interacting particles
confined in 1D and 2D. We consider hardcore bosons in spinless
Hubbard lattice models with short-range interactions.
We show that many-body states with topological
features emerge at different energy bands separated
by large gaps. The topology manifests in the way the particles
organize in real space to form states with different energy.
Each of these states contains topological defects/condensations
whose Euler characteristic can be used as
a topological number to categorize states belonging to the same energy band.
We provide analytical formulas
for this topological number and the full energy spectrum of the system
for both sparsely and densely filled systems.
Furthermore, we analyze the connection with the Gauss-Bonnet
theorem of differential geometry, by using the
curvature generated in real space by the particle structures.
Our result is a demonstration
of how states with topological characteristics, emerge in
strongly interacting many-body systems
following simple underlying rules, without considering the spin, long-range microscopic interactions, or external fields.
\end{abstract}

\maketitle 

\section{Introduction}
Self-organization mechanisms occuring in many-body systems,
that result in novel phases of matter, are one of the most important
pursuit in physics. One of the most celebrated examples is 
a universal mechanism of phase transitions, based on breaking of symmetries, 
invented by Landau\cite{Landau} and expanded by others, to describe long-range ordering phenomena like, ferromagnetism and solid, liquid and vapour 
transitions. Moreover, this universal mechanism
has been succesfully applied to describe phases formed under extreme conditions like, superconductivity and superfluidity. The key ingredient in Landau's theory is local
order parameters that take different values between the different phases
of matter, according to the relevant symmetries present. 
Later, other important self-organization
mechanisms related to topology were discovered,
that do not require long-range ordering, breaking of symmetries or order parameters
\cite{Berezinskii,Kosterlitz,Reppy,Ambegaokar,Agnolet,haldane0,wen,Gu,Levin,Chen}.
The resulting phases have been dubbed topological orders.
Apart from classical systems like superfluid
films\cite{Berezinskii,Kosterlitz,Reppy,Ambegaokar,Agnolet,Mathey,Scalettar}, topological orders can manifest also in the ground states of quantum systems with a large number of interacting particles, at low temperatures,
where quantum effects play an important role\cite{wen,Gu,Levin,Chen}.
Celebrated examples include quantum liquids like in the fractional quantum Hall effect
FQHE \cite{Tsui,Laughlin} and spin liquids\cite{Yoshida1,Isakov,Savary}. 
For such systems, topological order is strongly tied
to quantum correlations between the different components of the
system, known as entanglement\cite{amico,horodecki,Hamma,kim2,islam}.
In order to characterize topologically ordered phases, usually,
a global measure that takes into account the overall spatial properties of the system has to be used.
Widely used topological measures are for example, the winding number in the classical case\cite{Kosterlitz}, or the topological entanglement entropy\cite{Kitaev2} and the entanglement spectrum in the quantum case\cite{Li,Alba,Calabrese,Pollmann}. 
Topological order is an important
paradigm, showing that self-organization mechanisms leading to
states of matter, are not necessarily related to
long-range ordering and breaking of symmetries, 
as described in Landau's theory.

In this paper we study how strongly interacting
particles self-organize to create many-body states  
with topological characteristics in 1D and 2D,
based on simple microscopic rules.
We demonstrate this by using partially filled Hubbard models
with short-range interactions.
For strong interaction strength,
the particles organize in different sets of microstates.
We show that all microstates with the same energy
can be described by a topological number,
the Euler characteristic of the network/graph structures formed
by the empty or occupied space in the system.
We calculate geometrically this topological number 
along with the energy spectrum of the system,
for sparsely and densely filled systems, by using the topological
defects/condensations contained in the microstates.
Furthermore we discuss the connection with the Gauss-Bonnet theorem
of differential geometry, by using the curvature
generated by these topological structures.

\section{Model}
In order to demonstrate the topological self-organization mechanism
we consider spinless particles in 2D Hubbard lattices
or chains with short-range interactions, described by
the Hubbard-like Hamiltonian
\begin{equation}
\begin{aligned}
& H = H_U+H_t \\
& H_U = U \sum_{x=1}^{M_{x}}\sum_{y=1}^{M_{y}}(n_{x,y}n_{x+1,y} +  n_{x,y}n_{x,y+1}) \\
& H_t=t\sum_{x=1}^{M_{x}}\sum_{y=1}^{M_{y}}(c_{x+1,y}^{\dagger}c_{x,y} +  c_{x,y+1}^{\dagger}c_{x,y} + h.c. )\\
\end{aligned}
\label{eq_1}
\end{equation}
where $c_{xy}^{\dagger},c_{xy}$ are the creation and annihilation operators for spinless particles
at site with coordinates x,y in the lattice, while
$n_{x,y}=c_{x,y}^{\dagger}c_{x,y}$ is the number operator.
Also $M_{x}(M_{y})$ is the number of sites
along x(y) giving the total number of sites in the system $M=M_{x}M_{y}$.
Schematically, we represent occupied/unoccupied sites with
filled(empty) circles.
When two particles occupy adjacent sites in the lattice
they interact with energy U via the term $H_U$ in Eq. \ref{eq_1}.
This short-range interaction
leads to topological structures in the particle
arrangements in real space,  as we shall show
in the following section.
The interaction strength U
can be either positive or negative for repulsive
or attractive interaction, respectively.
All energies E in the paper are expressed in units
of U. A small nearest-neighbor hopping $t$
with $|U| \gg t$, described by $H_t$
can be treated perturbatively.
In order to avoid edge effects we close our system
in both directions x and y by applying periodic
boundary conditions (PBC) so that the summation indices in Eq. \ref{eq_1}
follow $M_x+1=1$ and $M_y+1=1$ giving $c_{M_{x}+1,y} = c_{1,y}$ and $c_{x,M_{y}+1} = c_{x,1}$.
For our study we consider hardcore bosons whose many-body wavefunctions stay symmetric
under exchange of two particles, but the particles cannot occupy the same quantum state.
In this case only one particle is allowed per site and $n_{i}$ can be either 0 or 1.
The hardcore bosons satisfy the commutation relation $[c_{i},c_{j}^{\dagger}] = (1-2n_{i})\delta_{ij}$.
They\cite{guo,wang1,zhang,Varney} can be realized in cold atom
and helium-4 systems experimentally~\cite{islam,goldman,Bloch}.
In overall our system corresponds to spinless hard-core bosons with strong
short-range interactions on the surface of a torus.
However our results could be extended to fermions
and other types of particles corresponding to different occupation
numbers, as the physical mechanism that generates the 
topological structures in the real space of the system is independent of the type of particles.

In order to characterize the topological structures
in our system, we use the Euler characteristic from graph theory and
differential geometry. It is defined as
\begin{equation}
\chi=V-\textit{E}
\label{eq_euler}
\end{equation}
where V is the number of vertices
and $\textit{E}$ the corresponding number of
edges between these vertices in the graph.

\section{Euler characteristic and energy spectrum}
At the strong interaction (Mott) limit $|U| \gg t$,
the particles localize at each
site of the Hubbard lattice. For $t=0$, by keeping
only the interaction term $H_U$ in Eq. \ref{eq_1}, the microstates
of the system and their corresponding energy are determined
by the way the particles organize in real space.
The energy of each microstate is determined by the number
of particle pairs, formed when two particles occupy adjacent sites,
or alternatively by the structure formed by the unoccupied sites (holes) in the lattice. The particles form what is known as charge density wave(CDW) states.
A first nearest-neighbor hopping described by $H_t$ in Eq. \ref{eq_1}
will cause strong mixing of the states for $t=0$.
However, other types of hoppings can preserve the states
 for $t=0$. We give some examples at the end of this chapter
 after analyzing the $t=0$ case.
Different particle configurations/microstates
can result in the same energy for $t=0$ or belong to the same energy band
when $t$ has a small non-zero value. In short, the
interaction between the particles splits
the Hilbert space of the non-interacting system (U=0)
in sub-spaces containing different many-body orders.
We have found that each of these sub-spaces can be characterized
by a topological number.
For dense systems, the empty space between the particles
forms 2D network/graph structures, whose Euler characteristic $\chi$
can be used as a topological number
to characterize the respective microstates.
In the rest of the paper we refer to these structures as topological defects,
which are also the driving mechanism in the Berezinskii-Kosterlitz-Thouless(BKT) transition\cite{Kosterlitz},
the prime example of topological order in classical systems.
The self-organization mechanism for $t=0$
is valid for both classical and quantum systems. 
However in the quantum case the system lies 
in a superposition of the microstates allowed at this energy.
This could be considered as a simplified version
of a quantum field, i.e., a coherent superposition
of states with point-particles arranged at different spatial configurations.

\begin{figure}
\begin{center}
\includegraphics[width=0.9\columnwidth,clip=true]{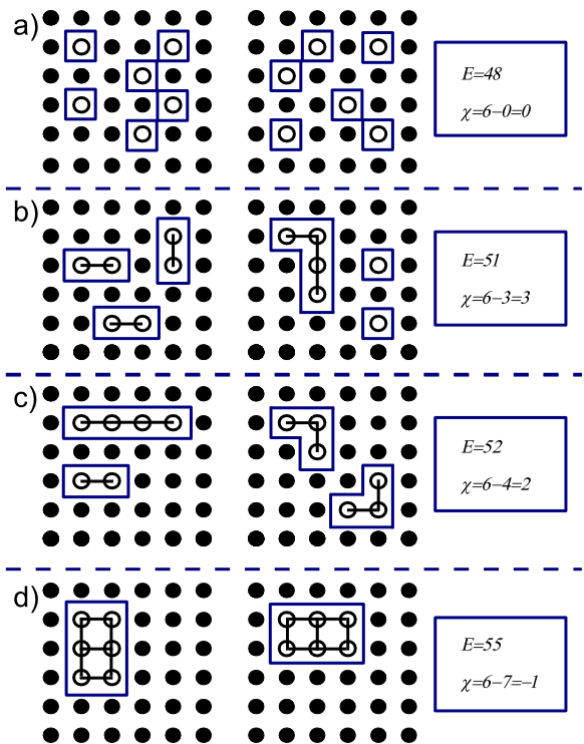}
\end{center}
\caption{Some particle configurations(microstates) for
a system with $M=36$ sites and $N=30$ particles, at different energies.
The empty space between the particles creates topological defects (blue boxes),
whose Euler characteristic $\chi$ is the same for microstates with the same energy.
a)All the defects consist of single unoccupied sites that are disconnected.
This gives the lowest energy of the system $E_{min}=48$,
geometrically corresponding to a graph with six vertices and no edges with $\chi_{max}=6$.
b) Three topological defects can be seen which can be deformed
maintaining the same energy as long as $\chi$ of their total structure
is preserved.
c) Same as b) but with two topological defects.
d)When all the unoccupied sites are connected the maximum energy of
the system is achieved $E_{max}=55$ which corresponds
geometrically to the minimum Euler $\chi_{min}=-1$
having the maximum edges between the vertices in the graph.
}
\label{fig_1}
\end{figure}

In the rest of the current section we analyze the 2D system which
concerns the major results of our paper.
In Fig. \ref{fig_1} we show a few examples of some microstates
containing different kinds of topological defects, along with
their corresponding energy and the Euler characteristic,
for a square system consisting of $M=36$ sites and $N=30$ particles.
We define $L=M-N$ as the number of unoccupied sites (holes).
The simplest application
of our idea is to notice that changing the position of the defects,
without deforming them, does not affect the total energy
of the system. A simple example of
this when the defects are single holes,
can be seen in Fig. \ref{fig_1}a.
Furthermore, the energy
does not change, if the defects are deformed
in a way that maintains the total Euler characteristic
of their structure (Eq. \ref{eq_euler}). This can be seen in Fig. \ref{fig_1}b,c for example.
Also, we notice that the number of defects does
not change for microstates with the same energy,
unless there are loops in the defects.
This already hints the topological character of the many-body orders,
as the defects could be considered as holes in a 2D
surface/manifold.

We can derive analytical formulas for dense systems
($L \ll N$) as follows.
For a dense system, the minimum energy of the system $E_{min}$ is achieved when all the unoccupied sites are disconnected. Since
each one of these holes has four nearest neighbors,
it reduces the total energy of the system by
\begin{equation}
E_{min}^{d}=4L.
\label{eq_2}
\end{equation}
In general the energy of each microstate
can be calculated by subtracting this defect energy $E^{d}$,
from the energy the system would have if all its sites were
occupied, that is, for a full system.
This energy is $E_{full}=(M_x-1)M_y+(M_y-1)M_x+(M_x+M_y)$
,where $M_{x}(M_{y})$ the number of sites
along x(y) with $M=M_{x}M_{y}$, giving
\begin{equation}
E_{full}=2M.
\label{eq_3}
\end{equation}
Then we can calculate the minimum(maximum) energy of the system from
\begin{equation}
E_{min(max)}=E_{full}-E_{min(max)}^{d}.
\label{eq_4}
\end{equation}
The corresponding  microstates which are the ground states
of the system, can be described
by an Euler characteristic when in geometrical terms
there are only $L$ unconnected vertices and no edges.
This is the maximum value of the Euler characteristic
\begin{equation}
\chi_{max}=L.
\label{eq_5}
\end{equation}
Also we observe that $E_{min}^{d}=4 \chi_{max}$.

As the individual defects/vertices become connected,
the energy of the system will be increased by a single step of U for every successive
set of microstates corresponding to the excited states.
On the other hand the corresponding Euler characteristic
will be decreased by one, as edges between
the vertices are added until the minimum Euler $\chi_{min}$ is reached
corresponding to the maximum energy $E_{max}$
of the system. This is achieved when all the holes
become connected by arranging in a lattice of parallelogram shape
as in Fig. 1d, for example. Any additional holes that do not fit in this parallelogram,
will be arranged in a way that maximizes the connections in the overall shape of the defect. This mechanism will result in a shape with the minimum number of holes with only one connection to the parallelogram (dangling bonds).
In geometrical terms this mechanism gives the maximum number of edges between the vertices. Using this idea we can derive analytical formulas for both
$E_{max}$ and $\chi_{min}$ as follows. We define
$L_{x}(L_{y})$ as the number of vertices
along x(y) with $L=L_{x}L_{y}$. Then the Euler characteristic
of this parallelogram lattice is $\chi=L-[(L_x-1)L_y+(L_y-1)L_x]=-L+(\frac{L}{L_y}+L_y)$
By minimizing this expression in respect to $L_y$ we get
\begin{equation}
\chi_{min}=Int[2\sqrt{L}-L].
\label{eq_6}
\end{equation}
Following a similar approach we can calculate
the maximum energy of the system by subtracting
from the total energy of the system the energy removed
by the defects $2L+(\frac{L}{L_y}+L_y)$.
Again by minimizing this expression in respect to $H_y$
we get $E_{max}^{d}=Int[2(\sqrt{L}+L)]$ which can
be expressed in terms of $\chi_{min}$ as
\begin{equation}
E_{max}^{d}=\chi_{min}+3L.
\label{eq_7}
\end{equation}
The maximum energy of the system is $E_{max}=E_{full}-E_{max}^{d}$.
In conclusion the system is split in different energy bands
determined by
\begin{equation}
E=E_{min},E_{min}+1,...,E_{max}
\label{eq_8}
\end{equation}
with corresponding Euler characteristic
\begin{equation}
\chi=\chi_{max},\chi_{max}-1,...,\chi_{min}.
\label{eq_9}
\end{equation}
with $E_{min},E_{max},\chi_{min},\chi_{max}$
given by Eq. (\ref{eq_2}-\ref{eq_7}).
Alternatively we can say that
the energy of the system is determined by $E=5N-3M+\chi$
in terms of $M,N,\chi$.

When the system is densely filled,
like the cases we considered, it is
reasonable to use its empty space
in order to define the Euler characteristic.
However, an alternative but equivalent
way to characterize the microstates would be to consider
the Euler characteristic of the structure
formed by the occupied sites in the Hubbard lattice.
This is more useful when examining a sparsely filled system
for $L \gg N$. In this case the minimum energy
of the system is $E_{min}=0$ when all the particles
are far apart from each other corresponding to $\chi_{max}=N$.
As the particles start occupying neighboring sites,
forming topological condensations,
the energy of the system will be increased by one step of U
for each set of microstates with higher energies.
The full energy spectrum
and the corresponding Euler numbers will be given
by Eq.\ref{eq_8} and Eq.\ref{eq_9} with
$E_{max}=Int[2(N-\sqrt{N})] $and $\chi_{min}=Int[2\sqrt{N}-N]$.

We remark that the topological defects
are equivalent to the structures
formed in a Hubbard model with attractive
interactions (-U) and $L$ particles
distributed in $M$ sites
(doing a particle-hole exchange in the original system).
This is an alternative way to generate
the topological structures that we have presented.

When the system is in a superposition
of states belonging to same energy band,
long-range spatial correlations
arise. In this state, all topological
defects are correlated with each other,
since any deformation of their patterns
should maintain the total Euler characteristic.
These correlations are independent of the
distance between the defects.
In this sense, the topological structures in our model
can be thought as patterns of quantum correlations,
which resemble the entanglement patterns in topologically
ordered quantum many-body systems.

We cannot smoothly deform microstates
with different Euler, between each other
without closing the energy gaps,
that is, without turning off the interaction between the
particles$(U=0)$.
Therefore, each set of microstates,
could be considered as a different topological state,
characterized by the Euler number
of the corresponding particle structures.

The analysis we presented so far is valid for vanishing
hopping term between the sites of the Hubbard lattice
($t=0$ in Eq. \ref{eq_1}). For finite nearest-neighbor
hopping a strong mixing between the states will occur
leading to different superpositions of the Fock states
than the ones we have analyzed. However, the microstates
for $t=0$ can be preserved by considering a hopping that respects
certain permutational symmetries satisfied by these states. 
For example, in order to preserve the internal
structure of the ground states, that contain single unconnected holes,
we can consider a hopping 
\begin{equation}
\begin{aligned}
 H_t=t\sum_{x=1}^{M_{x}}\sum_{y=1}^{M_{y}}(N_l c_{x+1,y}^{\dagger}c_{x,y} N_r + N_u c_{x,y+1}^{\dagger}c_{x,y} N_d + h.c.).
\end{aligned}
\label{eq_hop}
\end{equation}
$N_{r}=n_{x+2,y}n_{x+1,y+1}n_{x+1,y-1}$, $N_l=n_{x-1,y}n_{x,y+1}n_{x,y-1}$, $N_u=n_{x,y+2}n_{x+1,y+1}n_{x-1,y+1}$,$N_d=n_{x,y-1}n_{x+1,y}n_{x-1,y}$ check for
holes in the right, left, up and down direction, in respect to a hole lying
at coordinates $x,y$ in the lattice. 
This term prevents the clustering of the single holes in the ground state.
Similar hopping terms can be considered in order to preserve
the topological structures formed by the particles at higher energies. 
These hoppings should prevent the clustering of the corresponding defect structures
formed by the connected holes.

\section{Curvature}
The structures of the topological defects/condensations
generate a curvature at each site of the Hubbard lattice
that can be used to calculate
the Euler characteristic of each microstate.
If we use the defects, then the curvature at each site/vertice
can be defined as\cite{chen,oliver}
\begin{equation}
K(x,y)=1-\langle n_{x,y} \rangle-\frac{d(x,y)}{2}
\label{eq_10}
\end{equation}
where $\langle n_{x,y} \rangle$ is the occupation probability at
site with coordinates x,y in the Hubbard lattice.
The number of unoccupied neighboring sites d(x,y) is
\begin{equation}
d(x,y)=f(y_{+})+f(y_{-})+f(x_{+})+f(x_{-})
\label{eq_11}
\end{equation}
with
\begin{equation}
\begin{aligned}
& y_{+}=\langle n_{x,y} \rangle + \langle n_{x,y+1} \rangle \\
& y_{-}=\langle n_{x,y} \rangle + \langle n_{x,y-1} \rangle \\
& x_{+}=\langle n_{x,y} \rangle + \langle n_{x+1,y} \rangle \\
& x_{-}=\langle n_{x,y} \rangle + \langle n_{x-1,y} \rangle
\end{aligned}
\label{eq_12}
\end{equation}
and $f(v)=1-\textrm{H}(2v-1)$ where $\textrm{H}$ is the Heaviside step function
which obeys, $\textrm{H}(v)=0$ for $v<0$ and $\textrm{H}(v)=1$ for $v \geq 0$.
Notice that $K(x,y)=0$ for occupied sites
in the lattice($\langle n_{x,y} \rangle=1,f(v)=d(x,y)=0$)
and that the curvature takes only five values ($K(x,y)=-1,-1/2,0,1/2,1$) 
depending on the number of neighbors for each hole in the network.
The curvature can be interpreted as a local defect energy
at each site.
The Euler characteristic can be calculated
by summing the curvature over all sites of the Hubbard lattice
\begin{equation}
\chi=\sum^{x=M_x,y=M_y}_{x=1,y=1}K(x,y).
\label{eq_13}
\end{equation}
This is analogous to the integration of the curvature
of a closed geometrical shape (Euclidean manifold)
over its surface in the Gauss-Bonnet theorem
of differential geometry. The result
of this integration is always $2 \pi \chi$ with
$\chi=2-2g$ where $g$ is the genus counting the
number of holes in the geometrical shape.
Shapes with the same $g$ are topologically equivalent.
In our example, not all microstates with the same
$\chi$, belonging to the same energy band,
are topologically equivalent with each other.
This is due to the fact that there might exist states with
closed defects, containing loops which are
not topologically equivalent to open defect
structures. Therefore at each energy
there are subsets of microstates
that follow the same topology,
in the sense that the defects
contained in them, can be deformed
continuously between each other.

Nevertheless, still the overall structure of the system
in real space has to be taken into account,
in order to describe its physical properties.
This is a common characteristic of many topological phases
of matter.

In addition, since the system
contains quantum correlations between the
topological defects/condensations, an extension
of our approach, could provide insights into
the relation between entanglement and curvature
in many-body systems.

\section{1D system}
In the following we briefly analyze the
respective topological structures formed in a Hubbard
chain. In this case the characterization of the many-body states
is much easier to obtain since the particles form structures
only in a single direction.
As in the 2D case we
can use the Euler characteristic to describe
the structure of the topological defects formed by
the unoccupied sites in the lattice. The minimum
energy of the system occurs when all the unoccupied sites
are disconnected. It can be calculated by subtracting
the energy removed by the topological defects from the energy
of the system if all its sites were occupied. This gives
$E_{min}=2N-M$ for $N>M/2$ and $E_{min}=0$ for $N \le M/2$.
The corresponding maximum Euler of the topological
defects will be simply $\chi_{max}=M-N$. On the other hand
when all the unoccupied
sites or the particles become connected in a line, the maximum energy $E_{max}=N-1$
and minimum Euler $\chi_{min}=1$ are achieved.
In summary the 1D system can be described by Eq.\ref{eq_8} and Eq.\ref{eq_9},
as for the 2D system, but with $E_{min},E_{max},\chi_{min},\chi_{max}$
given above. The corresponding curvature at each site, whose
integration over the whole chain will give
the Euler number, can be easily obtained by
Eq. (\ref{eq_10}-\ref{eq_13}) after removing the terms
corresponding to the y direction $(y_{+},y_{-})$,
resulting only in two possible values K=0 or K=1.

\section{Summary and Conclusions}
We have shown how the self-organization of strongly interacting particles in 1D
and 2D, with simple underlying rules, can give rise to many-body states
with topological features. The topology
manifests in the structures formed by the particles
in real space, at different energies.
Consideration of the whole system is required to get
a complete understanding of its physical properties.
Each set of microstates belonging to the same
energy band can be described by a topological
number. For a dense system we have used
the Euler characteristic of the
defect structures formed by the empty space between the
particles. For a sparse system the corresponding Euler
of the condensations formed by the occupied space
in the system can be used instead.
A curvature is generated
by these topological defects/condensations that can be integrated over
the whole real space of the system to obtain the Euler
characteristic, as in the Gauss-Bonnet theorem
of differential geometry.
Our results show how
states with topological characteristics
emerge in partially filled many-body systems
with strong short-range interactions.

We hope that our findings motivate further
investigation of the physical mechanisms
that create states with topological features and 
other relevant emergent phases in many-body systems,
using simple microscopic rules.
Apart form their fundamental significance, such
mechanisms could be realized in cold atom
experiments, to design novel
phases of matter with various tunable properties, which
are a crucial step towards the realization of
quantum information applications.

\acknowledgments
We are thankful to Vladislav Popkov and Daw-Wei Wang for
useful comments.
We acknowledge resources and financial support provided by
the National Center of Theoretical Sciences in Hsinchu of R.O.C. Taiwan and the
Center for Theoretical Physics of Complex Systems in Daejeon of Korea under the project
IBS-R024-D1,

\end{document}